\documentstyle[aps]{revtex}

\begin{document}
\author{De-Zhong Cao$^1$, Jun Xiong$^1$, and Kaige Wang$^{2,1}\thanks{%
Corresponding author: wangkg@bnu.edu.cn}$}
\address{1.Department of Physics, Applied Optics Beijing Area Major Laboratory,\\
Beijing Normal University, Beijing 100875, China\\
2.CCAST (World Laboratory), P. O. Box 8730, Beijing 100080, China}
\title{Geometrical Optics in Coincidence Imaging System }
\maketitle

\begin{abstract}
We discuss the geometrical optics of coincidence imaging for two kinds of
spatial correlations which are related to a classical thermal light source and a
two-photon quantum entangled state.

PACS number(s): 42.50.Dv, 42.30.Va, 42.65.Lm
\end{abstract}

A novel imaging method, called coincidence imaging (or ghost imaging), has
drawn much attention recently\cite{shih1}-\cite{gatti}. In this imaging
system, object and image are separately illuminated by a pair of correlated
beams, and the image emerges with a coincidence detection of the two beams.
The first coincidence imaging experiments were carried out by a pair of
entangled photons generated in spontaneous parametric down-conversion (SPDC)%
\cite{shih1}\cite{rib1}. The effect can persist in a high gain of SPDC, in
which two entangled beams contain a large number of photons\cite{gigi}.
Recently, the experiments have shown that the coincidence imaging can be
emulated by classically correlated beams generated by randomly shooting of
rays\cite{ben1}. Theoretical studies have shown that a thermal source may
possess the similar spatial correlation as that of entangled photon pair\cite
{teich1}-\cite{gigi},\cite{gatti}-\cite{kaige2}. Therefore, the effects
related to two-photon entanglement, such as coincidence imaging, coincidence
interference and subwavelength lithography, may have classical counterparts.

In this paper, we focus on the study in the macroscopic aspect of
coincidence imaging: the geometrical optics. We find that the coincidence
imaging exhibits distinct aspect which cannot be included in ordinary
imaging. However, the difference of spatial correlation between quantum and
classical sources is also reflected in the geometrical optics. In the SPDC of
a type-I crystal, the down-converted beams contain both quantum entanglement
and classical thermal correlation. When the crystal is used as a source, it
can form a special dual coincidence imaging system, in which an object can
simultaneously produce two coincidence images. The system may find potential
application in new optical design.

We consider classical thermal light described by $E({\bf x},z,t)=\int E({\bf %
q})\exp [i{\bf q\cdot x]}d{\bf q\cdot }\exp [i(kz-\omega t)]$, in which $E(%
{\bf q})$ is a stochastic variable obeying Gaussian statistics and ${\bf q}$
is the transverse wavevector satisfying $|{\bf q}|<<k$. For any thermal
statistics, the second-order spectral correlation is written as 
\begin{eqnarray}
&&\langle E^{*}({\bf q}_1)E^{*}({\bf q}_2)E({\bf q}_2^{\prime })E({\bf q}%
_1^{\prime })\rangle  \label{6} \\
&=&\langle E^{*}({\bf q}_1)E({\bf q}_1^{\prime })\rangle \langle E^{*}({\bf q%
}_2)E({\bf q}_2^{\prime })\rangle +\langle E^{*}({\bf q}_1)E({\bf q}%
_2^{\prime })\rangle \langle E^{*}({\bf q}_2)E({\bf q}_1^{\prime })\rangle 
\nonumber \\
&=&S({\bf q}_1)S({\bf q}_2)[\delta ({\bf q}_1-{\bf q}_1^{\prime })\delta (%
{\bf q}_2-{\bf q}_2^{\prime })+\delta ({\bf q}_1-{\bf q}_2^{\prime })\delta (%
{\bf q}_2-{\bf q}_1^{\prime })],  \nonumber
\end{eqnarray}
where $S({\bf q})$ is the power spectrum of the spatial frequency. For
comparison, we show the second-order correlation of the entangled beams
generated in the lower gain limit of SPDC 
\begin{equation}
\langle a_m^{\dagger }({\bf q}_1)a_n^{\dagger }({\bf q}_2)a_n({\bf q}%
_2^{\prime })a_m({\bf q}_1^{\prime })\rangle =W^{*}({\bf q}_1)W({\bf q}%
_1^{\prime })\delta ({\bf q}_1+{\bf q}_2)\delta ({\bf q}_1^{\prime }+{\bf q}%
_2^{\prime }).  \label{3}
\end{equation}
where the subscripts $m$ and $n$ indicate the polarizations of the beams for
a type II crystal. The spectrum $W({\bf q})$ depends on the transfer
functions of SPDC\cite{kaige2}. The equation is also valid for a type I
crystal, in which the down-converted beams have the same polarization, and
the subscripts $m$ and $n$ can be omitted. Therefore, both the thermal light
and the entangled photon pair exist the transverse wavevector correlation.
The former shows the self-correlation of transverse wavevectors between
positive and negative components, while the latter shows the correlation of
a pair of conjugate wavevectors satisfying the momentum conservation, within
the same component of spatial frequency.

To show coincidence imaging of thermal light, we may use a 50/50
beamsplitter which divides input beam into two correlated beams. On the
other hand, in the collinear case of SPDC, the beamsplitter is also needed
to spatially separate the entangled down-converted beams. We define $F_i(%
{\bf q})$ ($i=1,2$) as the two output fields of the beamsplitter. For both
classical and quantum sources, the second-order correlation of the output
fields $\langle F_1^{*}({\bf q}_1)F_1({\bf q}_1^{\prime })F_2^{*}({\bf q}%
_2)F_2({\bf q}_2^{\prime })\rangle $ is proportional to that of the input,
i.e. Eqs. (\ref{6}) and (\ref{3}). For simplicity, we consider one dimension
case. Let $h_i(x,x^{\prime })$ ($i=1,2$) be the impulse response function
for the $i-$path in the coincidence imaging scheme, the joint-intensity at
the two detective planes is obtained to be 
\begin{equation}
\left\langle I_1(x_1)I_2(x_2)\right\rangle =\int
h_1^{*}(x_1,-q_1)h_1(x_1,-q_1^{\prime
})h_2^{*}(x_2,-q_2)h_2(x_2,-q_2^{\prime })\langle
F_1^{*}(q_1)F_1(q_1^{\prime })F_2^{*}(q_2)F_2(q_2^{\prime })\rangle
dq_1dq_1^{\prime }dq_2dq_2^{\prime },  \label{8}
\end{equation}
where $h_i(x,q)=(1/\sqrt{2\pi })\int h_i(x,x^{\prime })\exp (-iqx^{\prime
})dx^{\prime }$. Substituting Eqs. (\ref{6}) and (\ref{3}) into Eq. (\ref{8}%
), we obtain the joint-intensity 
\begin{mathletters}
\label{9}
\begin{eqnarray}
\left\langle I_1(x_1)I_2(x_2)\right\rangle &\propto &\int
S(q)|h_1(x_1,-q)|^2dq\int S(q)|h_2(x_2,-q)|^2dq+\left| \int
S(q)h_1^{*}(x_1,-q)h_2(x_2,-q)dq\right| ^2,  \label{9a} \\
\left\langle I_1(x_1)I_2(x_2)\right\rangle &\propto &\left| \int
W(q)h_1(x_1,-q)h_2(x_2,q)dq\right| ^2,  \label{9b}
\end{eqnarray}
for classical and quantum sources, respectively. Equations (\ref{9}) show
macroscopically the difference between classical and quantum coincidence
imaging. For the thermal source, the first term of Eq. (\ref{9a}) brings a
background while the second term devotes to the coincidence imaging.
Therefore, classical coincidence imaging has lower visibility than quantum
one. Furthermore, the nature of the wavevectors correlations for classical
and quantum sources is also reflected in the correlations of the two impulse
response functions. This will cause the different imaging laws.

Now we discuss the two schemes of coincidence imaging as shown in Fig. 1.
For simplicity, we assume that the beamsplitter is close to the source, so
that the beam is divided immediately from the source\cite{note}. For scheme
I, the two impulse response functions are written as 
\end{mathletters}
\begin{mathletters}
\label{10}
\begin{eqnarray}
h_1(x_1,q) &=&(1/\sqrt{2\pi })\exp [ikz_1-iqx_1-i\frac{q^2z_1}{2k}],
\label{10a} \\
h_2(x_2,q) &=&\frac 1{2\pi }\sqrt{\frac{kf}{i(f-z_3)f_c}}\exp
[ik(z_2+z_3+2f_c)-i\frac{q^2}{2k}(z_2+\frac{z_3f}{f-z_3})]  \label{10b} \\
&&\times \int T(x)\exp [i\frac{kx^2}{2(z_3-f)}-i(\frac{kx_2}{f_c}+\frac{qf}{%
f-z_3})x]dx,  \nonumber
\end{eqnarray}
where $f$ and $f_c$ are the focal lengths of the imaging lens F and the
collective lens F$_{\text{c}}$, respectively. $z_1$ and $z_2$ are the
distances from the source to detector D$_1$ and lens F, respectively; $z_3$
is the distance between object T and lens F. $T(x)$ is the transmission
function of object T. For scheme II, however, the two impulse response
functions are written as 
\end{mathletters}
\begin{mathletters}
\label{11}
\begin{eqnarray}
h_1(x_1,q) &=&\frac 1{2\pi }\sqrt{\frac k{if_c}}\exp [ik(z_1+2f_c)-i\frac{%
z_1q^2}{2k}]\int T(x)\exp [-i(\frac{kx_1}{f_c}+q)x]dx,  \label{11a} \\
h_2(x_2,q) &=&\sqrt{\frac f{2\pi (f-z_3)}}\exp [ik(z_2+z_3)-i\frac{q^2}{2k}%
(z_2+\frac{z_3f}{f-z_3})-i\frac{qx_2f}{f-z_3}-i\frac{kx_2^2}{2(f-z_3)}],
\label{11b}
\end{eqnarray}
where $z_1$ and $z_2$ are the distances from the source to object T and the
imaging lens F, respectively; $z_3$ is the distance between lens F and
detector D$_2$. In the broadband limit, $W(q)$ and $S(q)$ can be seen as a
constant in the integration, we calculate the joint-intensity by using Eq. (%
\ref{9}). For both the schemes, we introduce the coincidence imaging
equations 
\end{mathletters}
\begin{mathletters}
\label{12}
\begin{eqnarray}
\frac 1{z_2-z_1}+\frac 1{z_3} &=&\frac 1f,\qquad \text{for classical
coincidence imaging,}  \label{12a} \\
\frac 1{z_2+z_1}+\frac 1{z_3} &=&\frac 1f,\qquad \text{for quantum
coincidence imaging,}  \label{12b}
\end{eqnarray}
under which the coincidence imaging is obtained to be 
\end{mathletters}
\begin{equation}
\left\langle I_1(x_1)I_2(x_2)\right\rangle \sim 
{|T[x_1(f-z_3)/f]|^2,\qquad \text{for scheme I,} \atopwithdelims\{. |T[x_2f/(f-z_3)]|^2,\qquad \text{for scheme II.}}
\label{13}
\end{equation}
Note that Eq. (\ref{13}) is valid for both quantum and classical cases when
the background term is removed for the classical case. The coincidence
imaging (\ref{13}) is independent of position $x_2$ ($x_1$) of detector D$_2$
(D$_1$) for scheme I (II), since detector D$_2$ (D$_1$) and lens F$_{\text{c}%
}$ form a collective detection. However, $z_3$ is the object distance for
scheme I or the imaging distance for scheme II, so that Eq. (\ref{13}) gives
the same magnification as that in ordinary imaging.

In the coincidence imaging equation (\ref{12}), the joint-path $z_2\pm z_1$
is the imaging distance for scheme I or the object distance for scheme II,
reflecting the nature of the quantum and classical correlations. Though the
coincidence imaging equation is similar to the ordinary one, it will cause
rich and even surprising imaging effects that cannot be covered by the
ordinary imaging law, for example, a virtual image can become real, and vice
versa. Let us discuss two schemes in details.

{\em Scheme I}: When the object distance is greater than the focal length $%
z_3>f$, the joint-path $z_2\pm z_1$ as the image distance is positive. But
this does not assure a real coincidence image. Since $z_2$ is positive due
to definiteness, and $z_1$ could be either positive or negative: the former
causes a real coincidence image while the latter causes a virtual one. In
the case $z_3>f$, the condition for a real coincidence imaging is $%
(z_3-f)/(z_3f)-z_2>0$ ($<0$) for the source with quantum entanglement
(classical thermal correlation). Under the opposite condition, however, the
coincidence image is virtual. A virtual coincidence image can not be
directly observed in the coincidence detection.

Then we consider the case of the object distance less than the focal length $%
z_3<f$, for which the joint-path $z_2\pm z_1$ as the image distance is
negative. This derives that $z_1$ is negative for the quantum coincidence
imaging and positive for the classical coincidence imaging. Therefore, a
virtual image in the ordinary imaging system becomes real in the classical
coincidence imaging.

The coincidence imaging can be plotted by the graphics of the ray optics, by
taking into account the correlation of rays emitted from the source. In
graphics, we first plot the image by the ordinary way. Then, for the source
with quantum entanglement, the image is reflected twice, first by the source
and then by the beamsplitter. Obviously, the beamsplitter plays the role as
a mirror. However, the quantum source emits a pair of correlated rays with
the opposite transverse wavevectors, one to object and the other to image,
so that it also acts as a mirror. For the source with the thermal
correlation, the image is reflected only by the beamsplitter. This geometry
is due to the nature of the self-correlation of the wavevectors, so that the
thermal source acts as a phase-conjugate mirror and hence the image is
reflected to itself. According to these rules, we plot the coincidence
imaging for $z_3>f$ in Figs. 2 and 3, and $z_3<f$ in Fig. 4. For comparison,
we arrange the same optical setup for the two sources: the quantum
entanglement in Figs. 2a-4a and the classical thermal correlation in Figs
2b-4b. These figures verify the above analysis: while the quantum
coincidence image is virtual, the classical coincidence image must be real,
or vice versa.

{\em Scheme II}: In this scheme, the joint-path $z_2\pm z_1$ is the object
distance while $z_3$ is the image distance. Just as the ordinary imaging
law, when the joint-path $z_2\pm z_1$ is greater (less) than the focal
length, the coincidence image is real (virtual). Different from scheme I,
for the same optical setup with different sources, the quantum and classical
coincidence images can be both real. In Fig. 5, we plot the two real
coincidence images for $z_2\pm z_1>f$. In the graphics of this scheme, we
should move the object to the optical axes of the lens. For the source with
the quantum entanglement, the object is reflected twice, first by the
beamsplitter and then by the source, as shown in Fig. 5a. For the classical
correlation, however, the object is reflected only by the beamsplitter, as
shown in Fig. 5b.

The spatial thermal correlation (\ref{6}) exists in the SPDC process. In
type II SPDC, the quantum entanglement occurs between two beams with
different polarizations. But if one beam with a particular polarization is
extracted, it has the thermal correlation\cite{kaige2}. However, the beam
generated in type I SPDC may incorporate both the quantum entanglement and
the classical thermal correlation\cite{kaige2}. When the gain of SPDC is
lower, the power of the thermal correlation is lower than that of the
quantum entanglement, i.e. $|S(q)|<|W(q)|$. In the strong coupling of SPDC, $%
|S(q)|$ is increased and comparable with $|W(q)|$\cite{kaige}. Using this
source, it can form a dual coincidence imaging system, in which two kinds of
coincidence imaging are created simultaneously. For scheme I, classical
coincidence image is real while quantum one must be virtual, and vice versa.
For scheme II, however, two images can be both real, or both virtual, or one
is real and the other virtual, depending on the values of $z_1$ and $z_2$.

In summary, we show the macroscopic difference of quantum and classical
coincidence imaging. The unusual and rich coincidence imaging effects may
provide potential application in novel optical designs.

The authors thank L. A. Wu for helpful discussions. This research was
supported by the National Fundamental Research Program of China with No.
2001CB309310, and the National Natural Science Foundation of China, Project
Nos. 60278021 and 10074008.

Captions of Figures

Fig. 1 The sketches of coincidence imaging for (a) scheme I, where object T
and the imaging lens F are in the same path and (b) scheme II, where T and F
are in the different paths. F$_{\text{c}}$ is the collective lens so that
the object and the detector are placed in its two focal planes.

Fig. 2 Coincidence imaging for scheme I in which $z_3>f$ and the condition $%
\frac{z_3-f}{z_3f}-z_2>0$ are satisfied. (a) a real coincidence image is
formed for the source with the quantum entanglement; (b) a virtual
coincidence image is formed for the source with the thermal correlation. In
Figs. 2-5, the intermedial images are indicated by dashed lines.

Fig. 3 Same as in Fig. 2 but the condition $\frac{z_3-f}{z_3f}-z_2<0$ is
satisfied. (a) a virtual coincidence image is formed for the source with the
quantum entanglement; (b) a real coincidence image is formed for the source
with the thermal correlation.

Fig. 4 Coincidence imaging for scheme I in the case $z_3<f$. (a) a virtual
coincidence image is formed for the source with the quantum entanglement;
(b) a real coincidence image is formed for the source with the thermal
correlation.

Fig. 5 Coincidence imaging for scheme II in the case $z_2\pm z_1>f$. The
real coincidence images are formed for both the sources: (a) with the
quantum entanglement and (b) with the thermal correlation.


\begin{references}
\bibitem{note}  The further theoretical analysis shows that this assumption
is not necessary to obtain the imaging equations (\ref{12}).
\end{references}
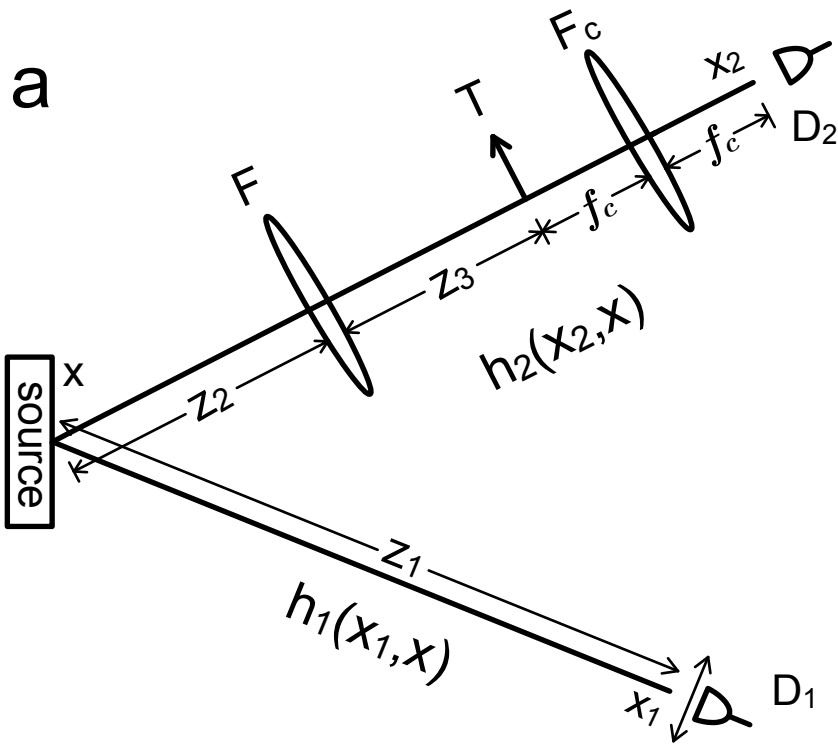
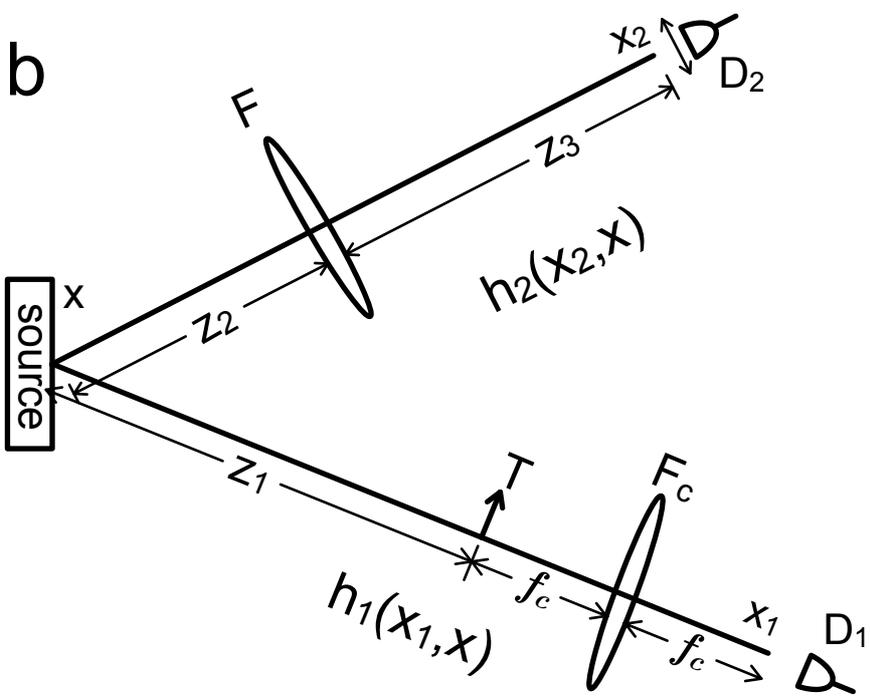
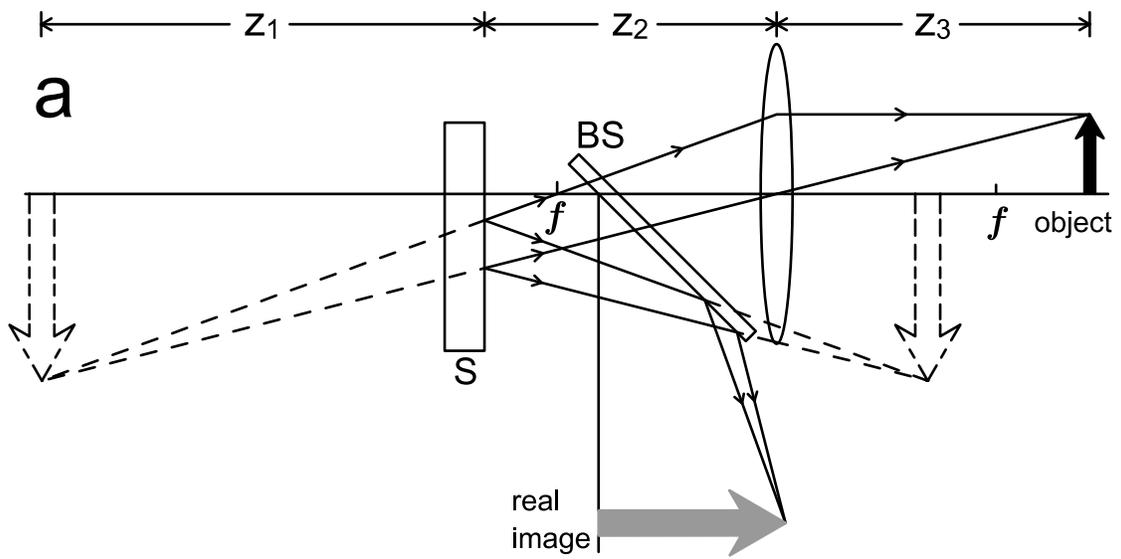
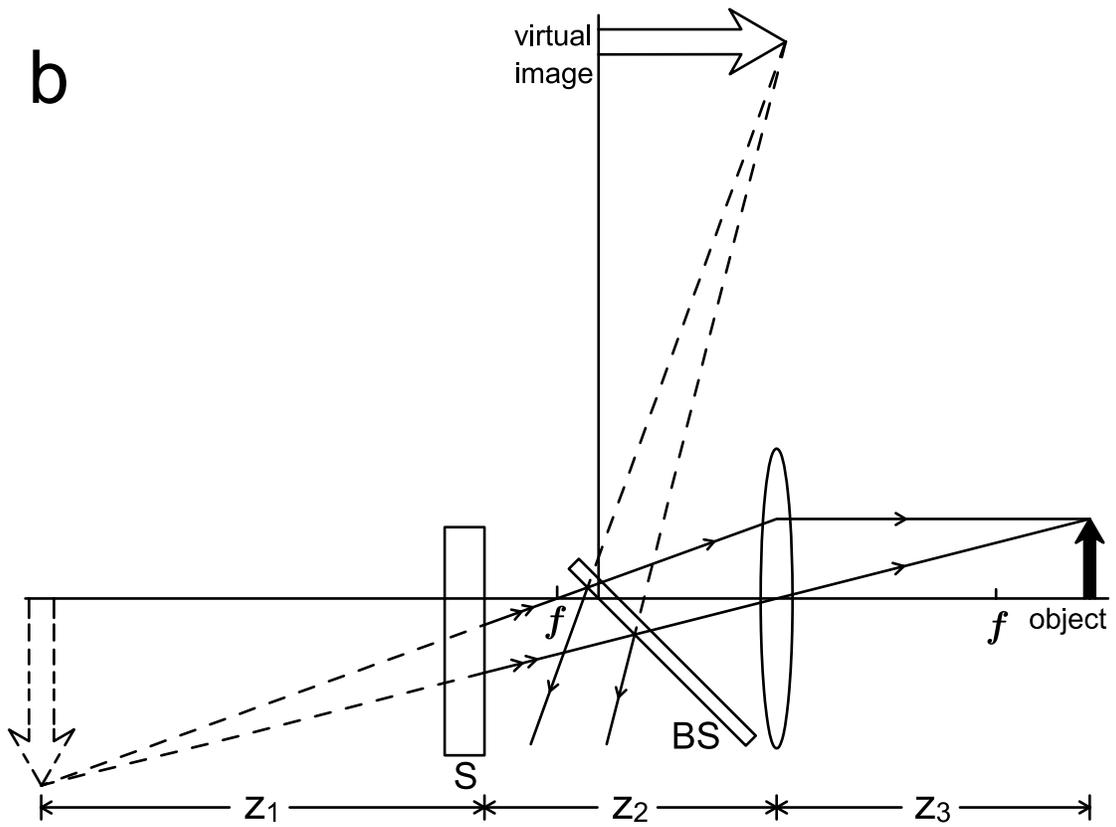
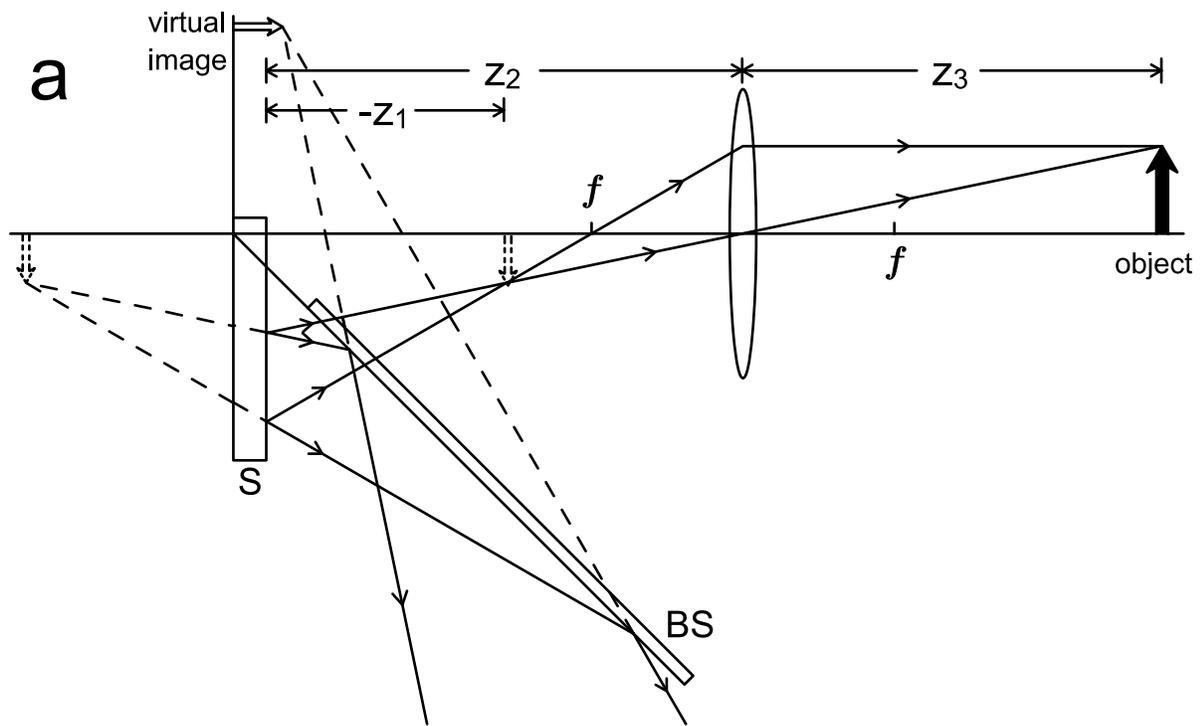
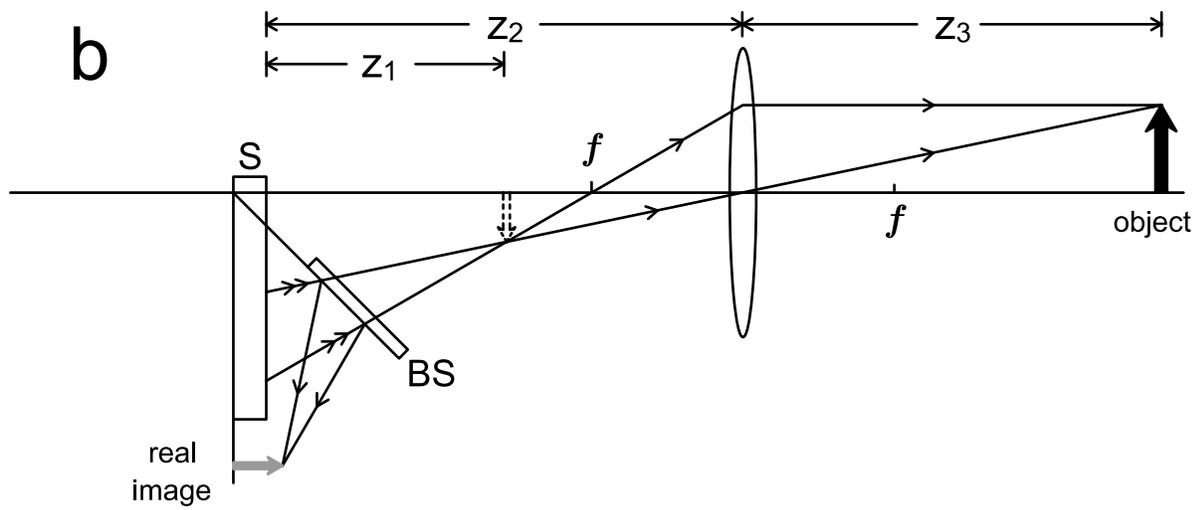
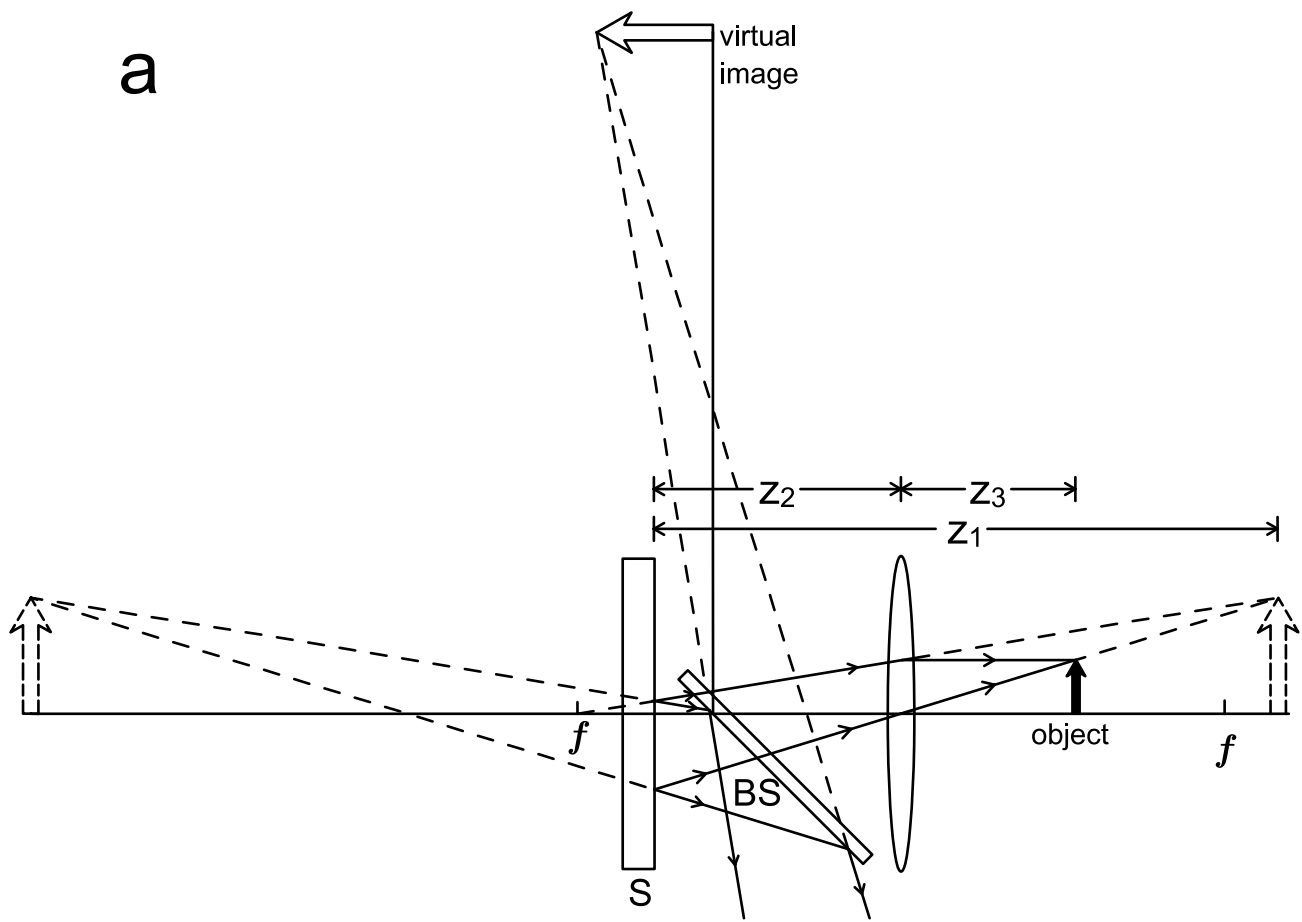
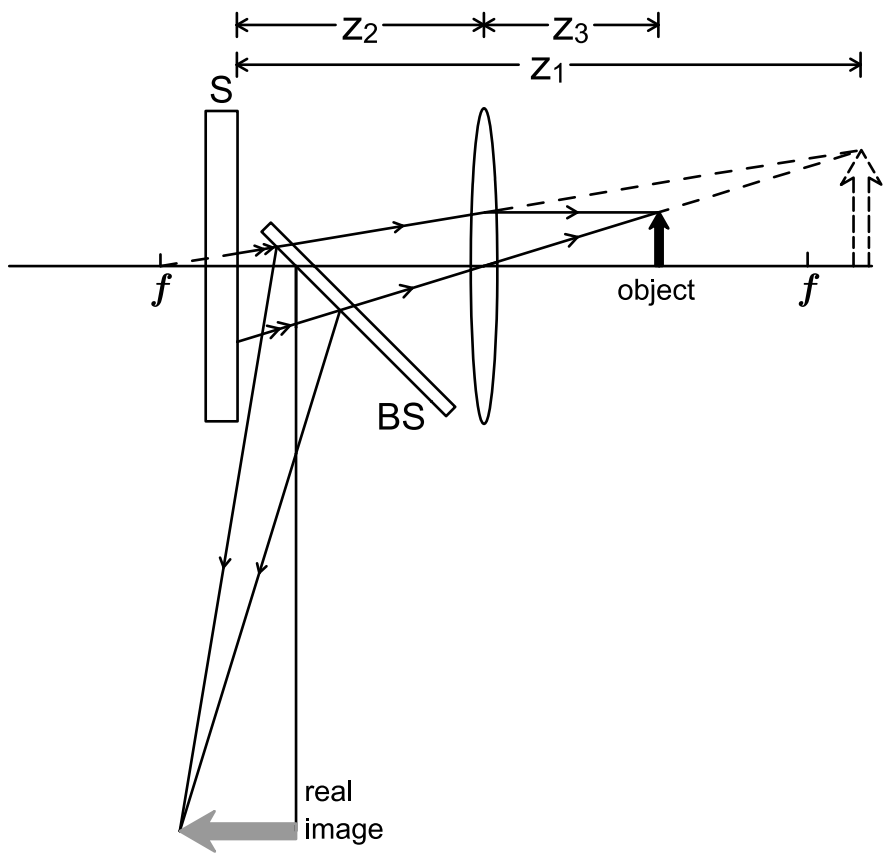
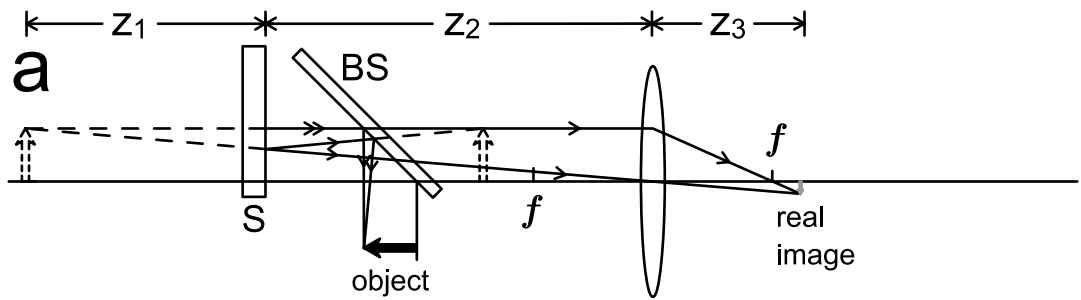
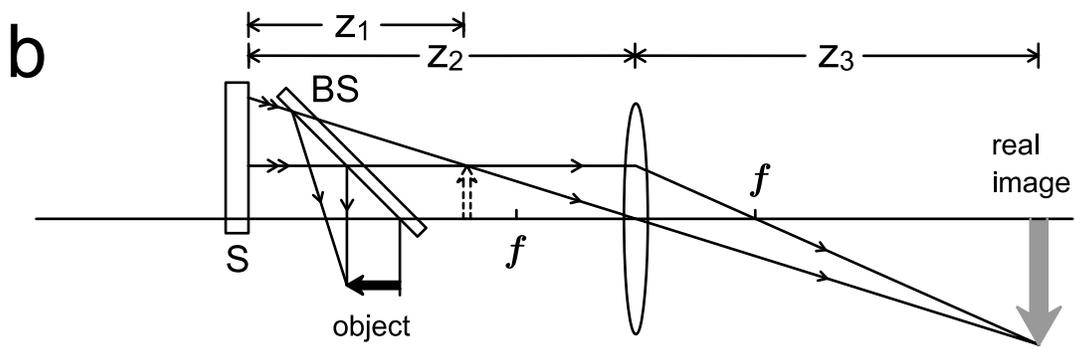
\end{document}